\documentstyle[preprint,aps,epsfig]{revtex}
\begin{document}
\draft
\title{Effects of geometric anisotropy 
 on local field distribution: Ewald-Kornfeld formulation}
\author{C. K. Lo, Jones T. K. Wan and K. W. Yu}
\address{Department of Physics, The Chinese University of Hong Kong, \\
         Shatin, New Territories, Hong Kong, China }
 
\maketitle

\begin{abstract}
We have applied the Ewald-Kornfeld formulation to a 
tetragonal lattice of point dipoles, in an attempt to examine 
the effects of geometric anisotropy on the local field distribution.
The various problems encountered in the computation of the conditionally
convergent summation of the near field are addressed and 
the methods of overcoming them are discussed.
The results show that the geometric anisotropy has a significant impact 
on the local field distribution.
The change in the local field can lead to a generalized Clausius-Mossotti
equation for the anisotropic case.
\end{abstract}
\vskip 5mm

\section{Introduction}

When a strong field is applied to a macroscopically heterogeneous or 
composite medium, the induced change of the medium can lead to quite 
interesting behavior, both in electrical transport and in optical response
\cite{ETOPIM5}.
These phenomena are most pronounced in the case of periodic composites. 
In the case of a cubic lattice of dipoles, elementary (dipole lattice) 
arguments show that the correction to the Lorentz cavity field 
(due to all the dipoles inside the cavity) strictly vanishes 
\cite{Ashcroft}. 

However, when the lattice symmetry is lowered by an external means, e.g., 
under the influence of an external force/torque, the lattice is deformed, 
either lengthened in one direction and/or contracted in the other 
direction, the correction to the cavity field will not vanish. 
We call this phenomenon the geometric anisotropy \cite{Yuen97}. 
The lattice deformation can easily be realized by the electrorheological 
effect on a suspension of polarized particles, in which the particles 
aggregate into anisotropic structures. 
In this work, we will apply the Ewald-Kornfeld formulation 
\cite{Ewald,Kornfeld} 
to a tetragonal lattice of point dipoles, in an attempt to obtain a
convergence of the infinite lattice sum and hence examine 
the effects of geometric anisotropy on the local field distribution.

The plan of the paper is as follows. In the next section, 
We will consider a tetragonal lattice of point dipoles and apply 
the Ewald-Kornfeld formulation to calculate the local field.
In section III, we perform numerical calculation and the various problems 
encountered in the computation of the conditionally
convergent summation of the near field will be addressed.
In section IV, we discuss the relation between the present formulation with
the established local field concepts.
We will show that the change in the local field can lead to a generalized 
Clausius-Mossotti equation valid for the anisotropic case.
Discussion on related problems will be given.

\section{Formalism}

Consider a tetragonal lattice with lattice constant $qa$ along the z-axis 
and lattice constant $aq^{-1/2}$ along the x and y axes. In this way,
the degree of anisotropy is measured by how $q$ is deviated from unity and
the uniaxial anisotropic axis is along the z-axis. 
The lattice constants have been chosen so that the volume of the unit cell
$V_c=a^3$ remains unchanged as $q$ varies. The lattice vector is given by
\begin{eqnarray}
{\bf R} = a(q^{-1/2} l\hat{\bf x} + q^{-1/2} m\hat{\bf y} + q n\hat{\bf z}),
\end{eqnarray}
where $l, m, n$ are integers. There are $N$ point dipoles ${\bf p}_i$ 
located at ${\bf r}_i$ in a unit cell. 
The local electric field ${\bf E}_i$ at a particular point dipole at 
${\bf r}_i$ can be expressed as a sum of the electric field of all dipoles 
at ${\bf r}_{{\bf R} j}$:
\begin{eqnarray}
{\bf E}_i = \sum'_j \sum_{\bf R} {\sf T}_{i {\bf R} j} \cdot 
  {\bf p}_j,
\label{dipole-sum}
\end{eqnarray}
where ``prime'' denotes a restricted summation which excludes $j=i$ when
${\bf R}=0$ and
\begin{eqnarray}
{\sf T}_{ij} = -\nabla_i \nabla_j {1\over |{\bf r}_i - {\bf r}_j|}
\end{eqnarray}
is the dipole interaction tensor.
Eq.(\ref{dipole-sum}) can be recast in the Ewald-Kornfeld form
\cite{Ewald,Kornfeld}:
\begin{eqnarray}
{\bf p}_i \cdot {\bf E}_i &=&
\sum'_j \sum_{\bf R} \left[-({\bf p}_i \cdot {\bf p}_j) B(r_{i {\bf R} j})
 + ({\bf p}_i \cdot {\bf r}_{i {\bf R} j})
   ({\bf p}_j \cdot {\bf r}_{i {\bf R} j}) C(r_{i {\bf R} j})\right] 
   \nonumber \\
 &-& {4\pi \over V_c} \sum_{{\bf G} \neq 0}
   {1\over G^2} \exp \left(-{G^2\over 4\eta^2} \right)
   [({\bf p}_i \cdot {\bf G}) \exp(i{\bf G} \cdot {\bf r}_i)
   \sum_j ({\bf p}_j \cdot {\bf G}) \exp(-i{\bf G} \cdot {\bf r}_j)]
   \nonumber \\
 &+& {4\eta^3 p_i^2 \over 3\sqrt{\pi}},
\label{E-K}
\end{eqnarray}
where $r_{i {\bf R} j}=|{\bf r}_i - {\bf r}_{{\bf R}j}|$, $\eta$ is 
an adjustable parameter and {\bf G} is a reciprocal lattice vector:
\begin{eqnarray}
{\bf G} = {2\pi\over a} (q^{1/2} u\hat{\bf x} + q^{1/2} v\hat{\bf y} 
 + q^{-1} w\hat{\bf z}).
\end{eqnarray}
The $B$ and $C$ coefficients are given by:
\begin{eqnarray}
B(r) &=& {{\rm erfc}(\eta r)\over r^3} + {2\eta \over \sqrt{\pi} r^2} 
  \exp(-\eta^2 r^2), \\
C(r) &=& {3 {\rm erfc}(\eta r)\over r^5} 
+\left({4\eta^3 \over \sqrt{\pi} r^2} + {6\eta \over \sqrt{\pi} r^4}\right)
   \exp(-\eta^2 r^2),
\end{eqnarray}
where ${\rm erfc}(r)$ is the complementary error function. 
Thus the dipole lattice sum of Eq.(\ref{dipole-sum}) becomes a summation 
over the real lattice vector {\bf R} as well as the reciprocal lattice 
vector {\bf G}.
Here we have considered an infinite lattice. For finite lattices, one 
must be careful about the effects of different boundary conditions
\cite{Allen}.
We should remark that although a tetragonal lattice is considered,
Eq.(\ref{E-K}) is applicable to arbitrary Bravais lattices.
The adjustable parameter $\eta$ is chosen so that both the summations 
in the real and reciprocal lattices converge most rapidly.
In what follows, we will limit ourselves to one dipole per unit cell, 
and the Ewald-Kornfeld summation [Eq.(\ref{E-K})] can be carried out.
We will consider two cases depending on whether the dipole moment is 
parallel or perpendicular to the uniaxial anisotropic axis. 
In both cases, we will compute the local field as a function of 
the degree of anisotropy $q$.

\section{Numerical Results}

Consider the longitudinal field case: ${\bf p}=p\hat{\bf z}$,
i.e., the dipole moments being along the uniaxial anisotropic axis. 
The local field {\bf E} at the the lattice point {\bf R} = 0 reduces to
\begin{eqnarray}
E_z = p\sum_{{\bf R}\neq 0}[-B(R)+{n^2}{q^2}C(R)] 
 - {4\pi p\over V_c} \sum_{{\bf G}\neq 0}{G_z^2\over G^2}
  \exp\left({-G^2\over 4\eta^2}\right) + {4p\eta^3\over {3\sqrt\pi}},
\end{eqnarray}
and $E_x=E_y=0$.
The local field will be computed by summing over all integer indices,
$(l,m,n) \neq (0,0,0)$ for the summation in the real lattice and 
$(u,v,w) \neq (0,0,0)$ for that in the reciprocal lattice. 
Because of the exponential factors, we may impose 
an upper limit to the indices, i.e., all indices ranging from $-L$ to $L$, 
where $L$ is a positive integer. 
For $q \neq 1$, the regions of summation will be rectangular rather than 
cubic in both the real and reciprocal lattices. 
The computation has been repeated for various degree of anisotropy with 
$q$ ranging from 0.5 to 2.0. A plateau value for $E_z$ is found for each 
$q$ within a certain range of $\eta$ values, indicating that convergence 
of the local field has indeed been achieved with the upper limit $L=4$. 

For a larger anisotropy, however, there have been two problems associated 
with the computation of the conditionally convergent summation:
(1) the range of $\eta$ that gives the plateau value shrinks as $q$ 
increases. Even no plateau value could be observed, say, for $q=0.1$.
In this case, the summation may still converge but possibly for 
a much larger $L$ and the computation time may be prohibitive.
(2) The local field can be used to evaluate the depolarization factor 
$\phi$, defined by: 
${\bf E}_{\rm far}=4\pi\phi{\bf P}={\bf E}_{\rm local} - {\bf E}_{\rm near}$,
where ${\bf P}={\bf p}/V_c$ is the total dipole moment per unit volume. 
The near field is the $\eta \to 0$ limiting value of the short-range part 
of the summation.
We find that if a direct summation over the near field is performed,
$\phi$ fluctuates seriously with the increase of $L$,
which is unacceptable.

In order to overcome these problems, the region of summation has been
taken to be inside a sphere of radius $Ja$ in the real lattice, 
and that of radius  $2\pi J/a$ in the reciprocal lattice, 
where $J$ is a positive integer. 
All the contribution from the dipoles outside the sphere will be discarded.
In this way, those dipoles that contribute significantly to the 
local field but were not considered in the rectangular box are
included in the summation.
As a result, the computation time can be shortened as a much smaller
value of $J$ can be used for convergence. 
The summation is repeated with increasing $J$ for convergence. 
We find that the summation indeed converges to a plateau value within 
a wide range of $\eta$ even for large anisotropy.

The second problem is also overcome by the summation over a sphere.
Although there are still some fluctuations of the $\phi$ value,
the amplitude of fluctuation is greatly reduced. We find that $\phi$
converges to 0.33 at $J=8$.
Physically, it reminds us that $\phi$ is exactly equal to $1/3$, 
independent of the degree of anisotropy. 
This implies that the far field is always equal to $4\pi{\bf P}/3$, 
as from far away, the lattice structure is irrelevant. 
What concerned us is just the total dipole moment per unit volume.
The fluctuation around $\phi=1/3$ is attributed to
the slow convergence and rapid fluctuation of the near field, 
rather than the local field which converges more rapidly.
This analysis thus provides us an accurate means of finding the near field  
by subtracting the far field, i.e., $4\pi{\bf P}/3$, from the local field.

For the transverse field case in which the dipole moments are perpendicular
to the uniaxial anisotropic axis, Eq.(8) can still be applied to evaluate
the local field by modifying $G_z$ to $G_x$ while taking the gradient 
along the direction of the dipole, say the x-axis, 
and obtain the expression of the local field.

The results of the local field strength (normalized to $4\pi P/3$) against 
$\log_{10} q$ for the longitudinal and transverse field cases are plotted 
in Fig.1(a) and Fig.1(b) respectively. For comparison,
the near field is also plotted on the same figure.
The near field vanishes at $q=1$, in accord with the previous result.
As $q$ decreases, the local field for the longitudinal field case increases 
rapidly while that for the transverse field case decreases rapidly.
In both cases, when $q$ deviates from unity,
the effect of geometric anisotropy has a pronounced effect on the
local field strengths.

\section{Contact with macroscopic concepts}

  \subsection{Generalized Clausius-Mossotti equation}

Our present theory is of microscopic origin, 
in the sense that we have computed the lattice summation by the 
Ewald-Kornfeld formulation. 
We have not invoked any macroscopic concepts like the Lorentz cavity field 
\cite{Bottcher1,Bottcher2,Lorentz} in the calculations.
However, to corroborate with these established concepts can lead to a 
modification of the Clausius-Mossotti equation valid for the anisotropic case. 

More precisely, we use the result of the local field to evaluate the 
effective polarizability $\alpha_{\rm eff}$ of the dipole lattice, 
which is given by :
\begin{eqnarray}
\alpha_{\rm eff}= {\alpha\over {1-\alpha\beta/V_c}},
\end{eqnarray}
where $\alpha$ is the polarizability of an isolated dipole, 
and $\beta=E/P$ is the local field factor.
Note that $\beta=4\pi/3$ when $q=1$.
To see this, the total field acting on a dipole is the sum of the 
applied field $E_0$ and the local field due to all other dipoles, hence
$$
p = \alpha (E_0 + \beta P),
$$
where $E_0$ is the applied electric field. 
Let $P=p/V_c$, the above equation becomes a self-consistent equation. 
Solving yields
$$
p= \left({\alpha\over {1-\alpha\beta/V_c}}\right) E_0.
$$
The effective dielectric constant $\epsilon_{\rm eff}$ is given by 
$1+4\pi\alpha_{\rm eff}/V_c$.
For a cubic lattice, $\beta=4\pi/3$, $\epsilon_{\rm eff}$ 
satisfies the well-known Clausius-Mossotti equation:
\begin{eqnarray}
{\epsilon_{\rm eff} - 1\over \epsilon_{\rm eff} + 2} = {4\pi \alpha \over 3V_c}.
\end{eqnarray}
Thus Eq.(9) represents a generalization of the Clausius-Mossotti equation
to the anisotropic lattice.

The result of the effective polarizability is plotted against 
${\log_{10}q}$ in Fig.1(c) for the longitudinal field case while in 
Fig.1(d) for the transverse field case with $\alpha=0.001, 0.01$ and $V_c=1$. 
For a small $\alpha=0.001$, the effective polarizability is almost 
independent of $q$.
However, for a larger $\alpha=0.01$, the effective polarizability exhibits 
similar behavior as the local field.
As $q$ decreases, $\alpha_{\rm eff}$ for the longitudinal field case increases 
rapidly while that for the transverse field case decreases rapidly.
Again, when $q$ deviates from unity, the effect of geometric anisotropy 
has a strong impact on the effective polarizability. 

  \subsection{Onsager reaction field}

A problem with the Lorentz theory is in regard to its generalization to 
polar media with permanent dipole moments $\mu$. A simple replacement of the 
polarizability $\alpha$ by $\alpha$ + $\mu^2/k_B T$ in the Clausius-Mossotti 
relation leads to a divergent dielectric constant, a phenomenon known as the
polarization catastrophy. In 1936, Onsager \cite{Onsager} resolved 
this problem by introducing a reaction field: 
while the Lorentz local field at the origin is due to all dipole moments 
of the lattice in the absence of the dipole moment at the origin, 
the reaction field at the origin arises from the additional polarization 
of the surrounding dipole moments due to the dipole moment ${\bf p}_0$ 
at the origin.

In the present case, the Onsager reaction field at the origin is given by:
\begin{eqnarray}
{\bf R}_0 = \alpha \sum_{{\bf R} \neq 0} 
 {\sf T}_{0{\bf R}} : {\sf T}_{{\bf R}0} \cdot {\bf p}_0,
\label{reaction}
\end{eqnarray}
where $\alpha$ is the bare polarizability and {\sf T} is the dipole 
interaction tensor
\begin{eqnarray}
{\sf T}_{0{\bf R}} = \nabla_{\bf R} \nabla_{\bf R} {1\over R} 
 = {\sf T}_{{\bf R}0}.
\end{eqnarray}
Unlike the summation of the local field, the infinite sum 
[Eq.(\ref{reaction})] 
for the Onsager reaction field is indeed absolutely convergent. 
No Ewald-Kornfeld formulation is needed because the product {\sf T}:{\sf T} 
is positive definite. 
For a simple cubic lattice of dipole moments, by summing over 
nearest neighbors, the result is:
$$
R_0 = 12\alpha p_0/V_c^2,
$$
which is already close to the infinite lattice limit
$R_0 \approx 16.8 \alpha p_0/V_c^2$.
It is instructive to extend the consideration to anisotropic lattices. 
The reaction field is conveniently expressed as 
$R_0=\lambda\alpha p_0/V_c^2$, where $\lambda$ is the reaction field factor.
Again, the summation over sphere has helped the convergence.
We perform the direct summation in the real lattice to obtain the reaction 
field factor for various $q$ ranging from 0.5 to 2.0.
In Fig.2(a) and (b), we plot $\lambda$ versus $\log_{10} q$ for the 
longitudinal and transverse field cases respectively. 
There is a minimum $\lambda$ around $q=1$ (but not exactly at $q=1$) 
in both cases and $\lambda$ increases rapidly when $q$ deviates from unity. 
As evident from Fig.2, the effect of geometry anisotropy has a strong 
impact on the reaction field strength.

\section{Discussion and conclusion}
Here a few comments on the results are in order.
We should remark that the present work employs the electrostatic 
(dipole) approximation. While such an approximation is simpler to 
implement and suffices in many cases, for optical properties of 
composites of metallic particles embedded in a dielectric host medium, 
however, one must go beyond the electrostatic approximation \cite{Yan}.

It is possible to extend the present theory to statistical geometric 
anisotropy \cite{Sheng91}, e.g., 
that being induced by the electrorheological effects \cite{Yuen97}.
We may also extend the formulation to a randomly dilute lattice of dipoles
to mimic mixed magnetic and nonmagnetic ions. 

Although we have employed a lattice structure in the present work, 
our formalism can readily be applied to an assembly of randomly 
placed dipoles by considering a more complex basis in a unit cell.
Such an extension will be useful for the study of dielectric liquids under
an intense electric field.

In conclusion, we have applied the Ewald-Kornfeld formulation to a 
tetragonal lattice of point dipoles to examine 
the effects of geometric anisotropy on the local field distribution.
The various problems encountered in the computation of the conditionally
convergent summation are addressed and the methods of overcoming them
are discussed.
We suggest that the large value of the derivative of the local field 
with respect to the degree of anisotropy offers potential applications
as artificial piezoelectric materials as one can change the 
degree of anisotropy easily in a suspension.

\section{Acknowledgments}
This work was supported by the Research Grants Council of the Hong Kong 
SAR Government under grant CUHK 4284/00P.

\begin{figure}[h]
\caption{(a) The normalized local field strength plotted against 
$\log_{10} q$ for 
dipole moments along the uniaxial anisotropic axis. 
When $q$ deviates from unity, the effect of geometric anisotropy has
a strong impact on the local field strengths.
(b) Similar to (a), but for dipole moments perpendicular to the uniaxial
anisotropic axis.
(c) The effective polarizability plotted against $\log_{10} q$ for 
dipole moments along the uniaxial anisotropic axis.
When $q$ deviates from unity, the effect of geometric anisotropy has
a similar effect on the effective polarizability.
(d) Similar to (c), but for dipole moments perpendicular to the uniaxial
anisotropic axis.}
\end{figure}
 
\begin{figure}[h]
\caption{(a) The reaction field factor plotted against $\log_{10} q$ for 
dipole moments along the uniaxial anisotropic axis. 
When $q$ deviates from unity, the effect of geometric anisotropy has
a strong impact on the reaction field strengths.
(b) Similar to (a), but for dipole moments perpendicular to the uniaxial
anisotropic axis.}
\end{figure}

\newpage
\centerline{Fig.1/Lo, Wan, Yu}
\centerline{\epsfig{file=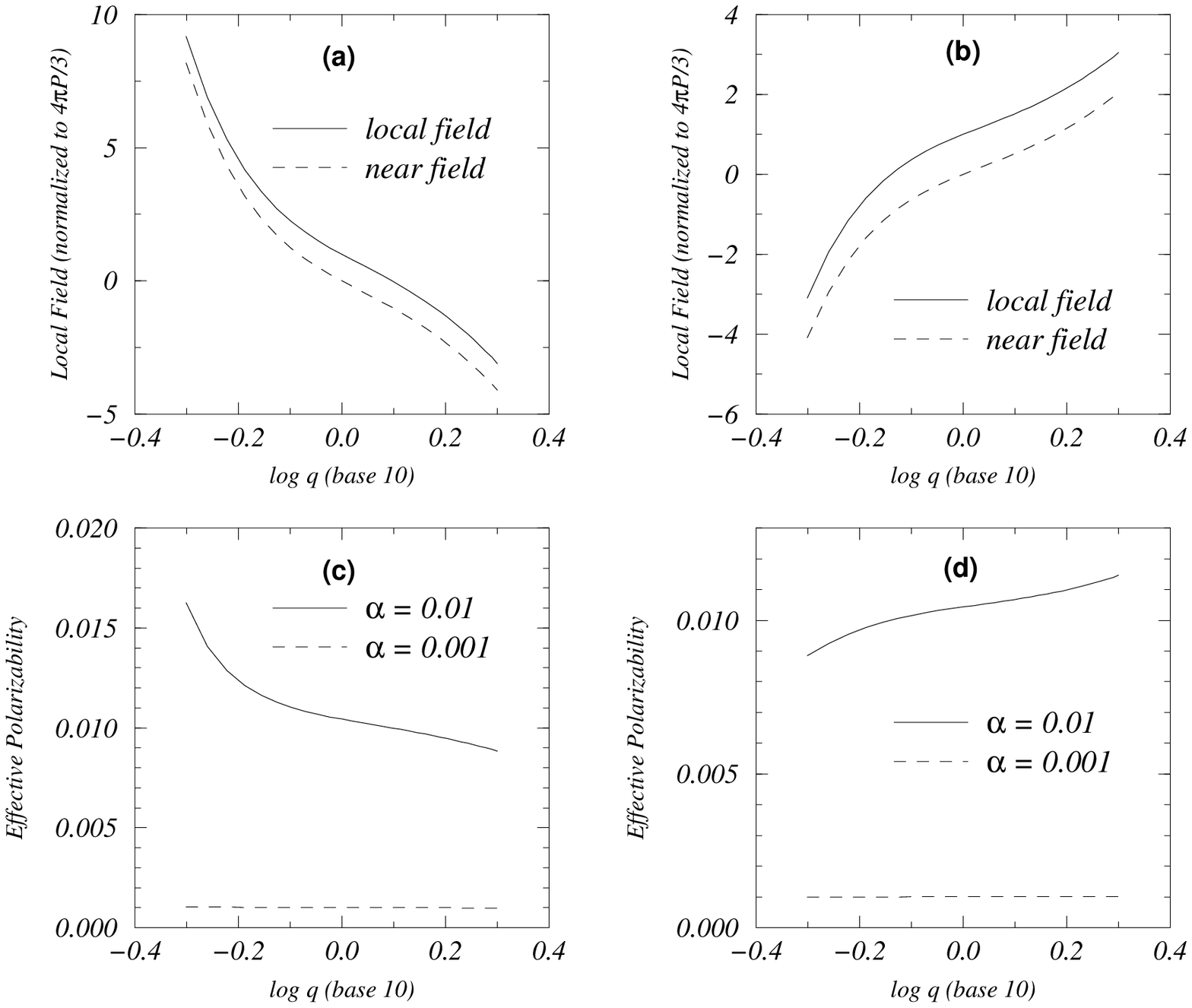,width=\linewidth}}

\centerline{Fig.2/Lo, Wan, Yu}
\centerline{\epsfig{file=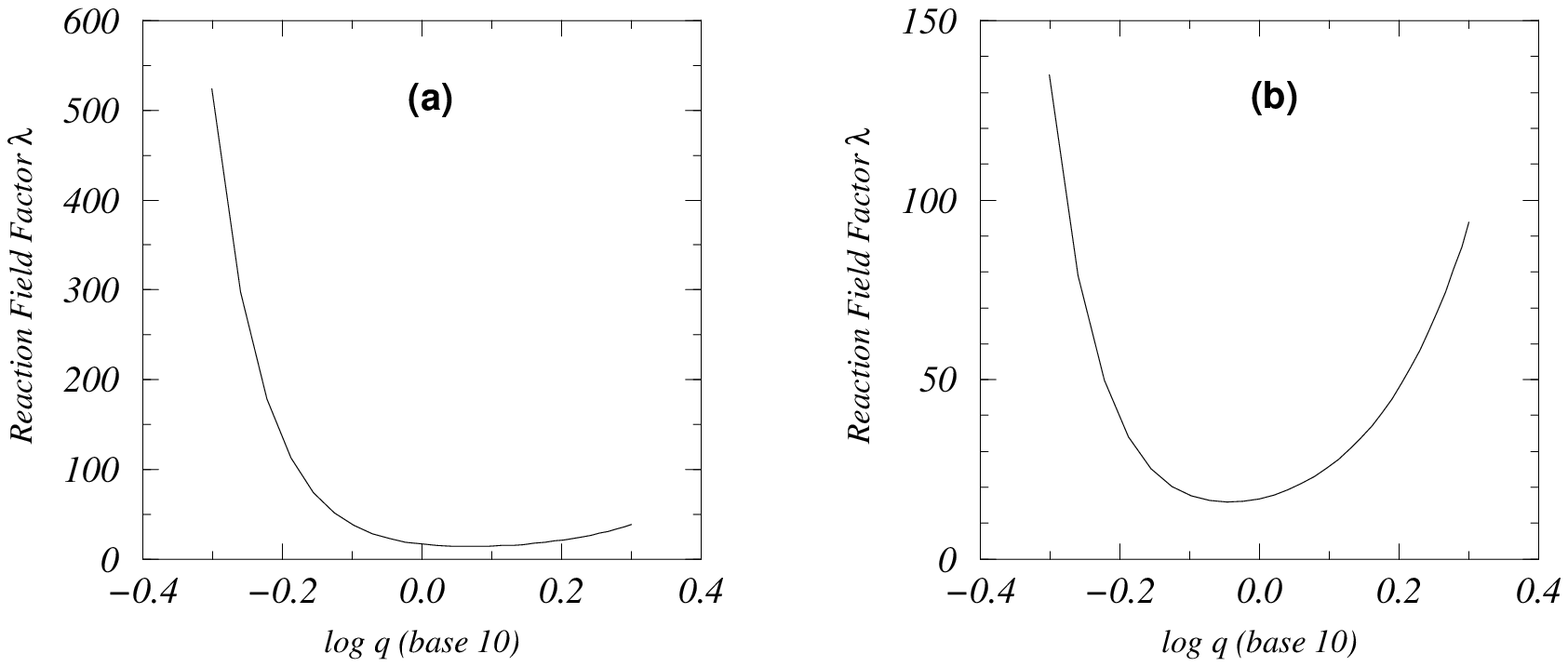,width=\linewidth}}


\begin{references}
\bibitem{ETOPIM5} For recent work, see {\em Proceedings of the 
 5th International Conference on Electrical Transport and Optical 
 Properties of Inhomogeneous Media}, Physica B {\bf 279}, (2000).
\bibitem{Ashcroft} N. W. Ashcroft and N. D. Mermin, 
 {\em Solid State Physics}, Chap. 27 (Holt, Rinehart and Winston, 1976);
 see also J. R. Reitz, F. J. Milford and R. W. Christy, {\em Foundations of
 Electromagnetic Theory}, Chap. 5, Third Edition, (Addison-Wesley, 1979).

\bibitem{Yuen97} K. P. Yuen, M. F. Law, K. W. Yu and Ping Sheng, 
 Phys. Rev. E {\bf 56}, R1322 (1997). 

\bibitem{Ewald} P. P. Ewald, Ann. Phys. (Leipzig) {\bf 64}, 253 (1921).
\bibitem{Kornfeld} H. Kornfeld, Z. Phys. {\bf 22}, 27 (1924);
 thesis, Goettingen (Germany), 1923.
\bibitem{Allen} M. Allen and D. Tildesley, {\em Computer Simulation of
 Liquids}, (Oxford Science, London, 1990).

\bibitem{Bottcher1} C. J. F. Bottcher, {\em Theory of Electric Polarization},
 Vol.I (Elsevier, Amsterdam, 1973).
\bibitem{Bottcher2} C. J. F. Bottcher and P. Bordewijk, 
 {\em Theory of Electric Polarization}, Vol.II (Elsevier, Amsterdam, 1978).

\bibitem{Lorentz} H. A. Lorentz, {\em The Theory of Electrons},
 (B. G. Teubner, Leipzig, 1909).
\bibitem{Onsager} L. Onsager, J. Am. Chem. Soc. {\bf 58}, 1486 (1936).

\bibitem{Yan} V. Yannopapas, A. Modinos and N. Stefanou, Phys. Rev. B
 {\bf 60}, 5359 (1999).
\bibitem{Sheng91} Z. Chen and Ping Sheng, Phys. Rev. B {\bf 43}, 
 5735 (1991).
\end{references}
\end{document}